\documentclass[multphys,vecphys]{svmult}
\usepackage{makeidx}     % allows index generation 
\usepackage{graphicx} 
\usepackage{multicol}    % used for the two-column index
\makeindex             % used for the subject index
\begin{document}

\title*{The Infrared Supernova Rate}
% Use \titlerunning{Short Title} for an abbreviated version of your
% contribution title if the original one is too long
\author{F. Mannucci\inst{1}
        G. Cresci\inst{2} 
        R. Maiolino\inst{3} 
        \and M.  Della Valle\inst{4}}
% Use \authorrunning{Short Title} for an abbreviated version of your
% contribution title if the original one is too long
\institute{ IRA-CNR, Largo E. Fermi 5, 50125 Firenze, Italy
\texttt{filippo@arcetri.astro.it} \and Dip. di Astronomia, Universit\'a
di Firenze, Largo E. Fermi 5, 50125 Firenze Italy
\texttt{gcresci@arcetri.astro.it} \and INAF-Osservatorio Astrofisico di
Arcetri, Largo E. Fermi 5, 50125 Firenze Italy
\texttt{maiolino@arcetri.astro.it} \and INAF-Osservatorio Astrofisico di
Arcetri, Largo E. Fermi 5, 50125 Firenze Italy
\texttt{massimo@arcetri.astro.it} }
%
% Use the package "url.sty" to avoid problems with special characters
\maketitle

\abstract
Optical searches can only detect supernovae (SNe) 
with a limited amount of dust extinction.
This is a severe limitation as most of the core-collapse SNe
could explode inside dusty regions. We describe a few ongoing projects aimed
at detecting dusty SNe at near-IR wavelengths both in ground-based and HST 
images and to study their properties.

\section{The problem} \label{sec:1}

Supernovae (SNe) exploding inside dusty regions could dominate,
even by a large amount,
the number of core-collapse events in the universe, 
as most of the star-forming activity is hidden by dust.
Nevertheless, centuries of optical searches have discovered only very few
SNe in dusty regions and no very obscured event. 

This is clearly a selection effect, and infrared or radio observations
are needed to reveal highly obscured SNe.  
Events detected at these long wavelengths can be used
to study the properties of the SNe in dusty galaxies,
to obtain a complete estimate of the total SN rate in the local universe, 
important
to calibrate the SN rate at high redshift now under study.
In principle the number of events could also be used to derive
information on the main energy source (starburst vs. AGN) 
of the galaxies when they are dominated by a hidden central source,
as for the Luminous Infrared Galaxies (LIRGS).

\section{Optical vs. near-infrared} \label{sec:1}

The observed rates of the core-collapse SNe, when derived from optical
observations and normalized to the B luminosity of the galaxy, don't
show any significant dependence on the galaxy type.  Normal galaxies
between Sa and Sm \cite{cap99}, starburst galaxies \cite{ric98},
galaxies with an active nuclei \cite{cap99}, and interacting galaxies
\cite{nav01}, all show the same SN rate of about 1 in SN units SNu
(number of SNe per
century per $10^{10}$ solar luminosities in the B band).

This is a puzzling result, as when a new episode of star formation 
starts in an old galaxy, both the B
luminosity and the SN rate increase (if the obscuration by the dust is
neglected) but the SN rate expressed in SNu 
is not expected to
remain constant. The SN rate, barely contaminated by the underlying
old population, is expected to show a sharper increase and evolve
on different
time scales. As a result the constancy of the SN rate cannot be explained
by dust-free models of galaxy evolution.

Large amount of dust are always present in starburst galaxies, effecting
both the B luminosity and the SN rate. Extinctions of $A_V\sim10$ are
often found, preventing the detection of SNe by optical observations.  It
is therefore crucial to use radio or infrared observation to derived a
more complete view of the SN events. Already in the near-infrared, at 2
$\mu$m of wavelength, dust extinction is much reduced, being about
1/10 of that in V.

When dealing with dusty active galaxies, the normalization based on the
B luminosity has no clear meaning as this band is produced by both the
old and new populations and is absorbed by the dust. In this case we
prefer to use the ``far infrared SN unit'' SNuIR, define as the number of SNe 
per century per $10^{10}$ solar luminosities in the Far InfraRed (FIR).
This normalization is more meaningful as the FIR luminosity 
is proportional to the current Star-Formation Rate
(SFR). It is actually possible to predict the number of expect SN from
the FIR luminosity \cite{man03,mat01}. This prediction depends on
several factors, as the radio properties of the SN 
(used
to estimate the intrinsic SN rate in nearby galaxies), the relation
between SFR and FIR luminosity, the Initial Mass function (IMF), the
presence of an AGN. The number of detected SN can also be used to
constrain these parameters.

\section{The ground-based observations} 
\label{sec:2}

Several groups have completed or started near-IR SN searches
\cite{van94,gro99,swirt,mat01}, but these works produced only two
detections and no spectroscopic follow-up. The reason of these negative
results are probably due to a combination of low spatial resolution, 
limited field-of-view, low sensitivity and small number of expected
events.

Our campaign started in 1999.
Observations up to 2001 are described in \cite{man03}, while in this
contribution we present the updated results up to summer 2003.
The galaxies were selected
to have large FIR luminosities, between $2\times10^{11} L_\odot$ and
$2\times10^{12} L_\odot$, corresponding to about 0.3$-$3 expected SNe 
per year per galaxy.  
Such high expected rates were chosen to assure significant
statistical results even in a short period of time.  The distances are
below 200 Mpc, assuring enough sensitivity and resolution to detect
point sources over the bright galaxy background.  We monitored 47
starburst galaxies in the K band (2.2$\mu$m)
mainly by using 4m class telescope in sites of good
seeing, the TNG in La Palma and the NTT at La Silla. Some observations
were also obtained by the UoA 61" telescope.
In 2002 and 2003 we obtained 50 new images, mainly with the NTT.
The total number of observations is now 304, with
an average number of 6.5 observations per galaxy.  
A sample of less
distant, less luminous galaxies were also monitored with the TIRGO 1.5m
telescope. The results will be discussed in a different paper (Cresci et
al., in preparation).

The various images of the same galaxy were carefully aligned, scaled to 
the same flux, reduced to the
same PSF and subtracted. Typical limiting magnitude were K$\sim$17
on the nucleus and K$\sim$19 at distances larger that about 1 arcsec.

These observations produced the detection of 4 events, the first
significant sample of events detected in the near-infrared. For one
event, SN2001db \cite{mai01}, we also obtain a spectroscopic follow-up:
this event is a type II SN discovered after maximum light. The
extinction, measured by the H$\beta$/H$\alpha$ and
Br$\alpha$/H$\alpha$ line ratios \cite{mai02} is $A_V\sim5.6$.
As expect, this one was the SN with the highest extinction know
at that time.

Obtaining an infrared SN rate from the data is not straightforward and is
subject to large uncertainties: this is due to the small number of
detected events, to the variability of the properties of core-collapse SNe
in the near-IR and to the dependence of the detection limit on the distance
from the galaxy nucleus. Using the same hypothesis of \cite{man03} we
derive an expected number of 55 SNe if they are all out of the nucleus
and 16 if they are in the central arcsec. Reducing these numbers to the
observed 4 events imply extinctions of $A_V=33$ and $A_V=10$,
respectively. The measured SN rate SN$_r^{NIR}$, 
assuming that 80\% of the SNe explode in
the nucleus (see \cite{man03}), is 0.40 SNuIR.

These limits are already quite high and are based on 4 years of (sparse)
observations. To do better it is necessary to use instruments with
higher sensitivity to point sources, as the HST.

\section{Archive HST observations} \label{sec:3}

The NICMOS camera on the HST is an ideal instrument to look for SNe in
the near-infrared. Its resolution is a few times higher than from the
ground under average seeing, and the PSF is much more stable allowing for
a better subtraction even near the bright galactic nucleus.

Most nearby starburst galaxies were already observed with this camera.
Unfortunately, the HST target selection policy does not usually 
allow for duplicate observations of the
same object with the same instrument setting. As a consequence, only
very few archive data can be used to look for variability. 

In a pilot study, we have searched the NICMOS archive for repeated
observations of starburst galaxies with a long time span.  We found 4
objects: NGC34, NGC5256, Arp200 and NGC6240.  For NGC34 and Arp220 a
narrow-band filter image was acquired a few months after the corresponding
broad-band one.  In this case each pixel of the broad-band image can be
"scaled" to the other bandpass by interpolation over the observed
broad-band colors.  
NGC5356 was observed twice with the same F160W filter but with a
different camera.  NGC6240 was observed twice in the same broad-band
filter and camera, probably because of problems with the PSF of the
first observation.  All the images have short exposure times, up to 4
minutes: nevertheless the resulting limit magnitudes are between H=18.8
and 21.0.  The nuclear region where the residuals of the subtraction are
large is confined in the central 0.3 arcsec, but we expect to reduce
it when the same instrument setting is used.

Despite the small sample of only 4 objects, the small number of
observations and the far than ideal instrument settings, 
the galaxies should produce about
5 observable SNe given their FIR luminosities,
the time span of the observations and limit magnitudes.
Despite of these expectations, no SN was detected. 
Also in this case we attribute this lack of detection
to the presence of high extinctions, but the
limits are less stringent.

\section{Incoming HST and VLA observations} \label{sec:4}

In order to obtain more useful data,
recently an HST proposal by our group was approved for cycle 12.
The aim of the proposal is to obtain second epoch
images of a sample of 37 nearby starburst galaxies already observed by NICMOS
in the F160W filter.  This ``snapshot'' program is already active and at
the moment of writing the first galaxy was already observed (see the figure).
If all the
galaxies will be observed, we expect to detect up to 50 SNe, value
corresponding to no extinction.  Even if all SNe suffer an extinction
of $A_V=30$ we would still expect to detect 8 SNe. 
Therefore we are looking forward these observations.  

\begin{figure} 
\includegraphics[height=4.8cm]{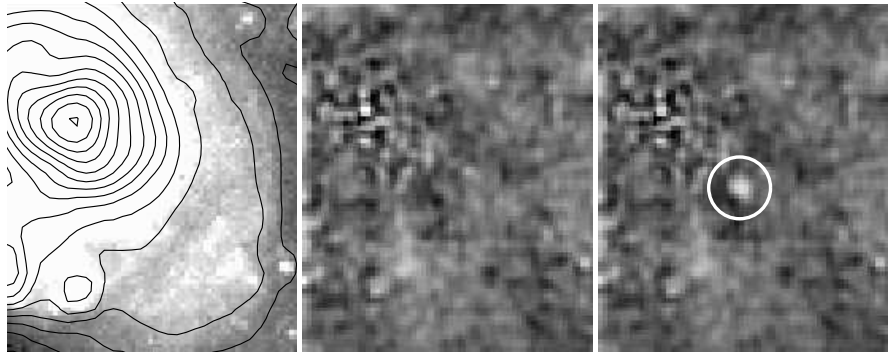}
\caption{ 
{\em Left panel:} NICMOS image of the nucleus of NGC3690 in the F160W 
filter taken in Aug 2003; 
{\em center panel:} residuals with an image taken in 1997 with the same
instrument setting; a fraction of less then 0.3\% of the flux in the
central arcsec (H$\sim13.6$) remains in the residual image.
{\em right panel:} a simulated SN with magnitude H=19.5 is added 
to show the residual noise level and the NICMOS detection power. 
The circle is 1 arcsec of diameter.
}
\end{figure}

All the detected objects will be observed spectroscopically. 
Any event detected in galaxies within 100 Mpc will be observed by VLA:
in these starburst galaxies we expect to find SNe with peculiar 
radio properties, as the emission
of the core-collapse SN at these wavelength is dominated by the interaction of 
the ejecta with the circumstellar medium.
The radio properties of these SNe can be used
to derive the density and the structure of the circumstellar medium and
the details of the late stages of the presupernova stellar evolution.
Comparison with the existing radio SN models will also test their
validity in a wider range of physical conditions than previously
available.

%%%%%%%%%%%%%%%%%%%%%%%% referenc.tex %%%%%%%%%%%%%%%%%%%%%%%%%%%%%%
% sample references
% "physics"
%
% Use this file as a template for your own input.
%
%%%%%%%%%%%%%%%%%%%%%%%% Springer-Verlag %%%%%%%%%%%%%%%%%%%%%%%%%%

%
% BibTeX users please use
% \bibliographystyle{}
% \bibliography{}

\printindex 
\end{document}